\documentclass[useAMS,usenatbib]{mn2e}
\usepackage{graphicx}
\usepackage{color}
\usepackage{amssymb}
\title[Pulsational behaviour of KUV 02464+3239]{Characterizing the pulsations of the ZZ Ceti star \mbox{KUV 02464+3239}}
\author[Zs. Bogn\'ar et al.]{
Zs. Bogn\'ar$^{1}$\thanks{E-mail: bognar[at]konkoly[dot]hu}, 
M. Papar\'o$^{1}$,
P. A. Bradley$^{2}$ and
A. Bischoff-Kim$^{3}$
\\
$^{1}$Konkoly Observatory of the Hungarian Academy of Sciences, P.O. Box 67., H--1525 Budapest, Hungary\\
$^{2}$X-4, MS T-087, Los Alamos National Laboratory, Los Alamos, NM 87545, USA\\
$^{3}$Department of Chemistry, Physics and Astronomy, Georgia College \& State University, CBX 82, Milledgeville, GA 31061, USA\\
}

\begin{document}

\date{Accepted 2009 ..... Received 2009 ...; in original form 2009 ...}

\pagerange{\pageref{firstpage}--\pageref{lastpage}} \pubyear{2009}

\maketitle

\label{firstpage}

%%%%%%%%%%%%%%%%%%%%%%%%%%%%%%

\begin{abstract}
We present the results on period search and modeling of the cool DAV star KUV 02464+3239. Our observations resolved the multiperiodic pulsational behaviour of the star. In agreement with its position near the red edge of the DAV instability strip, it shows large amplitude, long period pulsation modes, and has a strongly non-sinusoidal light curve. We determined 6 frequencies as normal modes and revealed remarkable short-term amplitude variations. A rigorous test was performed for the possible source of amplitude variation: beating of modes, effect of noise, unresolved frequencies or rotational triplets. Among the best-fit models resulting from a grid search, we selected 3 that gave $l=1$ solutions for the largest amplitude modes. These models had masses of 0.645, 0.650 and 0.680\,$M_{\sun}$. The 3 `favoured' models have $M_{\rmn{H}}$ between $2.5\,\rmn{x}\,10^{-5} - 6.3\,\rmn{x}\,10^{-6}\,M_*$ and give 14.2 -- 14.8\,mas seismological parallax. The 0.645\,$M_{\sun}$ (11\,400\,K) model also matches the spectroscopic $\rmn{log\,} g$ and $T_{\rmn{eff}}$ within 1\,$\sigma$. We investigated the possibility of mode trapping and concluded that while it can explain high amplitude modes, it is not required. 
\end{abstract}


%%%%%%%%%%%%%%%%%%%%%%%%%%%%%%

\section{Introduction}

Since Arlo Landolt's (1968) discovery of the first DAV \mbox{(HL Tau 76)} we have learned much about pulsating white dwarfs. They are otherwise normal white dwarf stars, and they represent the final stage of about 98\% of the stars (see the recent review of \citealt{winget2} and references therein), so it is a key issue to gain insight into their interiors. While white dwarfs are relatively simple to model, some observed phenomena still challenge theory.

One unsolved mystery is the variability of the pulsational modes' amplitudes and frequencies of PNNV, cool DBV and cool DAV stars \citep{handler1}. While we can find very stable modes in certain white dwarfs which can be used to measure evolutionary effects (cooling and/or contraction, e.g. hot DAVs G117-B15A and R548 -- `The Most Stable Optical Clocks Known': Mukadam, Kepler \& Winget 2001 and references therein), in some cases short-term variabilities make it difficult to determine even the pulsational modes. An example for this latter is the case of PG 1115+158 (DBV) with its remarkably unstable amplitude spectra \citep{handler2}.

In some cases we encounter strange, sudden, short-term variations in the pulsational behaviours of stars: for example the `sforzando' effect observed in the DBV star GD 358 in 1996, where the nonlinear light curve of the star changed into a high amplitude, remarkably sinusoidal one for a short period of time (\citealt{kepler1}, \citealt{provencal1}). In the case of PG 1456+103 its luminosity variations almost disappeared just before the Whole Earth Telescope (WET, \citealt{nather1}) run started on the star \citep{handler1}. The frequency and amplitude variations of multiplet components (mostly for high $k$ modes) observed in GD 358 \citep{provencal1} are further examples for short-term variations.

Possible explanations for these observed phenomena could be non-linear mode coupling, convection/pulsation interactions, mode excitation, and beating of different pulsation modes -- such as unresolved rotational splitting. To find the right explanation(s) for a certain case is a great challenge. A crucial first step is to identify the periods of the true pulsation modes of a white dwarf.

One possible way of mode identification when strong amplitude variations occur is to observe the star in many seasons, determine all the observed periods from the different runs and search for the (nearly) equidistant period spacings among them, which can be consecutive overtones with the same horizontal degree ($l$) value. \citet{kleinman1} used this method for the first time for a DAV star (G29-38) where recurring sets of normal modes were identified.

Since KUV 02464+3239 was found to be a luminosity variable DA star near the red edge of the DAV instability strip \citep{fontaine1}, we expect to find the short-term variabilities which characterize the pulsation of similar stars. Using a 2900\,s-long light curve, \citet{fontaine1} determined a quasi-period of $\sim$832\,s for the star's light variation. The long period, large amplitude variations they found with strongly non-sinusoidal light curve along with the presence of harmonic peaks in the frequency spectrum are consistent with the behaviour of a cool DAV star. However, up to now only the results of this short discovery run had been published about this target.

In this paper we present the results of observational analysis of KUV 02464+3239. We describe our observations and data reduction process in Sect.~\ref{obs}. The results of Fourier analyses and frequency determination tests can be found in Sect.~\ref{fourier}. Sect.~\ref{Ampl.var.} contains our tests on amplitude variations. We discuss the results of asteroseismological investigations (modeling, results for stellar parameters) in Sect.~\ref{seism} and give a brief summary in Sect.~\ref{sum}.

%%%%%%%%%%%%%%%%%%%%%%%%%%%%%%

\section[]{Observations and data reduction}
\label{obs}

We observed KUV 02464+3239 on 20 nights between October 2006 and February 2007. The observations were made with a Princeton Instruments VersArray:1300B back-illuminated CCD camera\footnote{http://www.princetoninstruments.com/products/imcam/\\versarray/dsheet.aspx} attached to the 1m RCC telescope at Piszk\'estet\H o mountain station of Konkoly Observatory. The log of observations is in Table~\ref{obslog}. We did not use any filter (`white light observations') to maximize the photon count from this relatively faint target ($m_v = \rmn{16\fm07}$). The back-illuminated detector we used is more sensitive to bluer wavelengths than a front-illuminated one. Note that the amplitude values of the Fourier light curve solution corresponding to these white light measurements differ from filtered ones.

Standard IRAF\footnote{IRAF is distributed by the National Optical Astronomy Observatories, which are operated by the Association of Universities for Research in Astronomy, Inc., under cooperative agreement with the National Science Foundation.} routines were used during the reduction process; the frames were bias, dark and flat corrected before we performed aperture photometry with the IRAF DAOPHOT package. The observational times of every data points were calculated to Barycentric Julian Date (BJD).

Different apertures were used for every night. For a given night, we chose the aperture size that resulted in the lowest scatter for differential light curves of constant stars in the field.

We used 10 or 30\,s exposure times depending on the weather conditions during the observing runs. For 10\,s exposure we averaged the data into 30\,s bins to keep all the data with the same bin, although, we are aware of that for multiperiodic pulsators with nearby pulsations it does introduce an extra uncertainty.

\begin{table}
\caption{Journal of observations of KUV 02464+3239. Subset No. refers to subsets in Sect.~\ref{Analysis.ref.interval.}, $N$ denotes the number of points remained after averaging (see details in text) and $\delta T$ means the time span of the runs including gaps.}
\label{obslog}
\begin{center}
\begin{tabular}{p{4mm}p{6mm}ccrr}
\hline
Run & Subset & Date & Start time & $N$ & $\delta T$\\
No. & No. & (UT) & (BJD-2\,450\,000) & & (h)\\
\hline
01 & 1 & 2006 Oct 06 & 4014.577 & 210 & 2.02\\
02 & 1 & 2006 Oct 07 & 4015.584 & 117 & 1.31\\
03 & 1 & 2006 Oct 09 & 4017.546 & 216 & 2.37\\
04 & 1 & 2006 Oct 11 & 4019.543 & 264 & 2.98\\
05 & & 2006 Oct 25 & 4034.283 & 304 & 6.29\\
06 & & 2006 Nov 26 & 4065.603 & 112 & 1.27\\
07 & & 2006 Nov 27 & 4066.542 & 65 & 0.87\\
08 & 2 & 2006 Nov 28 & 4068.206 & 1163 & 11.17\\
09 & & 2006 Dec 07 & 4077.299 & 278 & 4.11\\
10 & & 2006 Dec 08 & 4078.474 & 182 & 1.89\\
11 & 3 & 2006 Dec 11 & 4081.169 & 732 & 8.72\\
12 & & 2006 Dec 12 & 4082.435 & 81 & 0.75\\
13 & & 2006 Dec 13 & 4083.176 & 477 & 9.92\\
14 & 4 & 2006 Dec 14 & 4084.170 & 1048 & 10.92\\
15 & 4 & 2006 Dec 15 & 4085.179 & 877 & 10.56\\
16 & 4 & 2006 Dec 16 & 4086.187 & 696 & 7.54\\
17 & & 2006 Dec 17 & 4087.322 & 107 & 1.14\\
18 & 4 & 2006 Dec 19 & 4089.208 & 747 & 6.90\\
19 & & 2007 Jan 29 & 4130.216 & 362 & 5.16\\
20 & & 2007 Feb 19 & 4151.289 & 189 & 1.91\\
\hline
Total: & & & & 8227 & 97.80\\
\hline
\end{tabular}
\end{center}
\end{table}

\begin{figure}
\begin{center}
\includegraphics[width=6cm]{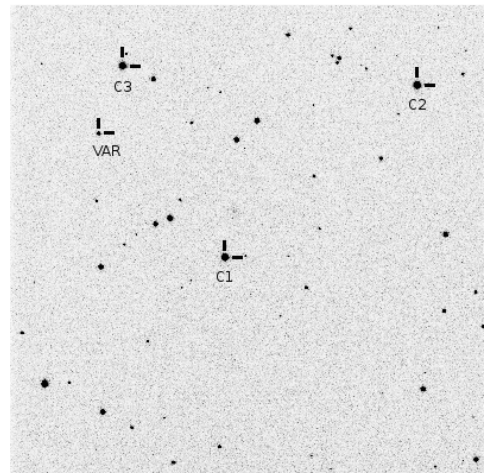}
\end{center}
\caption{Variable and comparison stars in the CCD field. The three brightest stars (C1, C2, C3) were selected as a reference system for the differential photometry.}
\label{map}
\end{figure}

\begin{figure}
\begin{center}
\includegraphics[width=8cm]{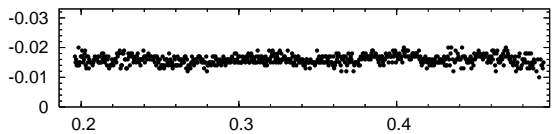}
\end{center}
\caption{Differential light curve of the check star (C3) to the average of the reference stars obtained on JD 2\,454\,086.}
\label{checkstar}
\end{figure}

\begin{figure*}
\begin{center}
\includegraphics[angle=0,width=17.5cm]{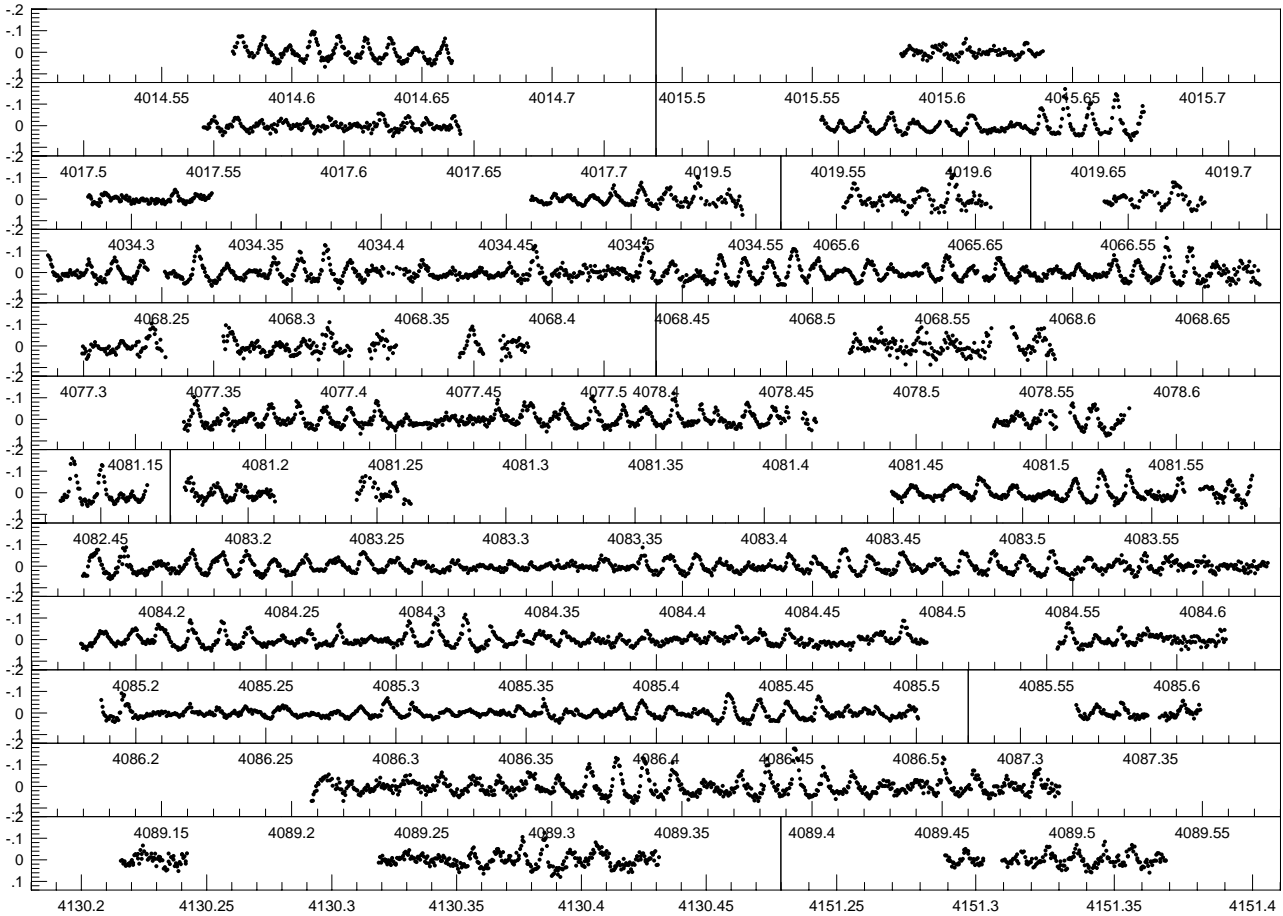}
\end{center}
\caption{Normalized differential light curves of the 20 runs obtained on KUV 02464+3239 between October 2006 and February 2007. Time is given as BJD-2\,450\,000.}
\label{lightcurves}
\end{figure*}

We performed aperture photometry on the same 22 stars for every night. Possible comparison stars were checked for variability and longer-term trends (caused by colour effects of stars with different spectral types) in an iterative way. This process resulted in our choosing the three brightest stars marked in Fig.~\ref{map} as a reference system for the differential photometry. The quality of this method is shown by the differential light curve of a check star (C3) compared to the average of the reference stars in Fig.~\ref{checkstar}.

Following the traditional method of photoelectric photometry, we obtained first-order extinction coefficients for the nights which were long enough. The values lay in the range of 0.07--0.39 in white light. 

The colour term of extinction was checked. However, we could not precisely determine the second order extinction coefficient for each night. In addition to the iterative process, we obtained \textit{BVRI} photometry to estimate the spectral types of the comparison stars selected in the iterative process. The relative colour indices were checked. The differences between the \textit{B-V} colour indices are higher than $\sim$0.5 which shows immediately that the comparison stars have much later spectral types than KUV 02464+3239. Naturally, this deficiency often resulted in trends in the differential light curves and accordingly significant signals in the low frequency region of their Fourier Transform (FT). Although the low frequency signals do not exceed 1--6\,mmag, to get a homogeneous dataset for the variable star, polynomial fits were performed in the last step of the reduction process. It is widely applied in white dwarf research using multichannel photometry. The light curves obtained for each run can be seen in Fig.~\ref{lightcurves}.

\section[]{Fourier analysis}
\label{fourier}

Since the star was not well studied before, our first aim was to find the frequency content.
Analyses were made by use of the MuFrAn (Multi-Frequency Analyzer) package \citep{kollath1, csubry1}. MuFrAn provides efficient tools for frequency determination with its standard analyzer applications (FFT, DFT, linear and nonlinear fitting options) and graphic display routines. The program handles unequally spaced and gapped observational data. Frequency values are given in cycle/day\,(c/d) units.

We followed the standard steps of a pre-whitening process to get frequency, amplitude and phase values utilising the light curve's Fourier spectrum. In pre-whitening processes the question is always when to stop iterating. In deciding whether a peak belongs to a real pulsation frequency, we used the widely accepted approximation: if a peak reaches the threshold of signal-to-noise ratio of $\sim$4 it has a small probability of being due to noise \citep{breger1}. 

The tools of the time string analysis program Period04 \citep{lenz1} were used to calculate S/N of the individual peaks. The program determines the signal (S) value as the amplitude of the peak after the least-squares fit. The noise (N) is given as an average amplitude calculated in a frequency range that encloses the peak in the residual spectrum. We determined S/N values from the $\pm$25\,c/d frequency range of peaks after pre-whitening the original FT.

\subsection[]{Investigation of individual nights}

\begin{figure}
\begin{center}
\includegraphics[width=9.0cm]{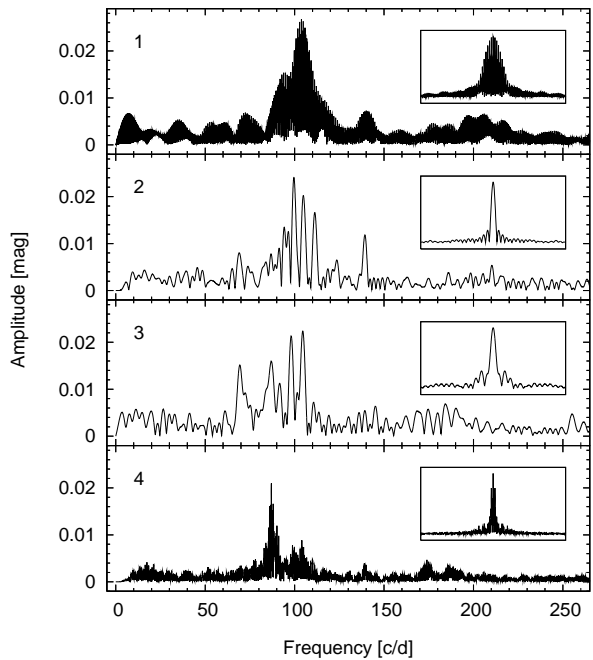}
\end{center}
\caption{Amplitude spectra of the subsets. Numbers of subsets are indicated in the left-upper corners of each panels. Window functions are given in the inserts. Remarkable amplitude variations can be seen from one interval to another.}
\label{FT.ref.nights}
\end{figure}

\begin{table}
\caption{Frequencies detected with S/N\,$\geq$\,4 values in the Fourier Transforms of subsets. Period and amplitude values are given for completeness.}
\label{ref.nights.freq}
\begin{center}
\begin{tabular}{rrrr}
\hline
\multicolumn{1}{c}{Frequency} & \multicolumn{1}{c}{Period} & \multicolumn{1}{c}{Ampl.} & \multicolumn{1}{c}{S/N}\\
\multicolumn{1}{c}{(c/d)} & \multicolumn{1}{c}{(s)} & \multicolumn{1}{c}{(mmag)} & \\
\hline
\multicolumn{4}{l}{Subset 1}\\

90.660$\pm$0.004 & 953.01 & 12.8 & 4.9\\
99.251$\pm$0.003 & 870.52 & 17.4 & 6.6\\
103.790$\pm$0.002 & 832.45 & 30.6 & 11.9\\
114.103$\pm$0.004 & 757.21 & 10.0 & 4.0\\[1.5mm]

\multicolumn{4}{l}{Subset 2}\\

99.567$\pm$0.028 & 867.76 & 23.6 & 7.1\\
104.880$\pm$0.036 & 823.80 & 20.2 & 6.5\\
111.452$\pm$0.045 & 775.22 & 15.1 & 5.6\\
139.227$\pm$0.063 & 620.57 & 11.6 & 6.0\\[1.5mm]

\multicolumn{4}{l}{Subset 3}\\

69.475$\pm$0.047 & 1243.61 & 14.7 & 5.4\\
87.264$\pm$0.044  & 990.09 & 16.3 & 6.0\\
98.025$\pm$0.053  & 881.41 & 16.1 & 6.0\\
104.358$\pm$0.047 & 827.92 & 18.0 & 7.3\\[1.5mm]

\multicolumn{4}{l}{Subset 4}\\

$^*$86.579$\pm$0.004 & 997.93 & 9.5 & 5.5\\
86.996$\pm$0.002 & 993.15 & 18.4 & 10.6\\
$^*$89.142$\pm$0.003 & 969.24 & 10.7 & 6.2\\
$^*$90.100$\pm$0.001 & 958.94 & 15.2 & 8.8\\
98.797$\pm$0.002 & 874.52 & 9.1 & 5.5\\
104.036$\pm$0.002 & 830.48 & 10.2 & 6.5\\
139.954$\pm$0.005 & 617.35 & 3.8 & 4.7\\
173.960$\pm$0.004 & 496.66 & 4.7 & 4.3\\
\hline
\end{tabular}
\end{center}
$^*$Closely spaced frequencies to the dominant peak. Signals at 89.14 and 90.10\,c/d are close to the 1\,c/d aliases of each others. The peak at 90.1\,c/d has an unreliably high amplitude solution only when we fit at least the first six frequencies of this subset.
\end{table}

Because the pulsation periods are short compared to the typical run length, the one-night long observing runs provide a long enough time base to investigate the basic pulsational characteristic of the star, even in a case if not all the modes are resolved. In addition, a night by night analysis allows us to follow the changes in pulsation amplitudes.

Considering the threshold of S/N$\sim$4 for the peaks, our results show that the pulsation frequencies of KUV 02464+3239 are at $\sim$87, 97, 104, 111 and 139\,c/d. In one case we also find a very significant (S/N=5.4) signal at $\sim$70\,c/d. During the pre-whitening process we did not take into account peaks under 60\,c/d in the FTs. This limit is an overestimation of the frequency range filtered by polynomial fitting and this means that any possible pulsation frequency below this level remained undiscovered.

Our analysis revealed that remarkable amplitude variations can happen from one night to another as in many cases of PNNV, cool DBV and cool DAV stars (\citealt{handler1}).

\subsection[]{Analysis of reference intervals}
\label{Analysis.ref.interval.}

The dataset can be grouped into four subsets of selected nights. The longer time base means better spectral resolution and with the proper selection of nights, the time base still remains short enough not to obscure the possible short-term variabilities. 

The first four nights (JD 2\,454\,014\,--\,2\,454\,019) were selected from October (subset No. 1). We got the longest time string during the eighth run (JD 2\,454\,068) so we used this run as a reference for November (subset No. 2). In December we had the chance to observe KUV 02464+3239 in two consecutive weeks. In view of the results of individual nights' Fourier analyses we selected two intervals: the first is the run on JD 2\,454\,081 (subset No. 3), the second consists of four nights (subset No. 4; JD 2\,454\,084\,--\,2\,454\,089, except the short run on JD 2\,454\,087). The latter runs have similar FTs but the run on JD 2\,454\,081 has a different FT.

Fig.~\ref{FT.ref.nights} shows the FTs of our subsets. The variations in amplitudes of peaks in the FTs are obvious. Frequency values determined after pre-whitening are given in Table~\ref{ref.nights.freq}. S/N values of peaks were also calculated. Noise levels change from lower to higher frequencies between 2.6-2.5\,mmag (subset 1), 3.3-1.9\,mmag (subset 2), 2.7-2.5\,mmag (subset 3) and 1.7-1\,mmag (subset 4). Standard deviations of frequency values are determined by Monte Carlo (MC) simulations, getting solutions for each frequency in each subset. We generated synthetic light curves (by adding Gaussian random noise) corresponding to the frequencies can be seen in Table~\ref{ref.nights.freq}. After the non-linear least squares fitting of 100 synthetic datasets of each subset we determined the standard deviations of the frequencies.
%The scatter of the noise added was determined from the noise statistics of our CCD measurements taking into account the brightness of the variable.

\begin{table}
\caption{Frequency and amplitude values of the 6 accepted pulsation frequencies for each subset. Amplitudes were determined by fitting with these frequency components only. Subset No. refers to subsets of Sect.~\ref{Analysis.ref.interval.}}
\label{ref.nights.freq2}
\begin{center}
\begin{tabular}{lcccccc}
\hline
Subset & \multicolumn{6}{c}{Frequency}\\
No. & \multicolumn{6}{c}{(c/d)}\\
\hline
1 & & & 99.25 & 103.79 & & \\
2 & & & 99.57 & 104.88 & 111.45 & 139.23\\
3 & 69.48 & 87.26 & 98.02 & 104.36 & &\\
4 & & 86.99 & 98.80 & 104.04 & &\\
\hline
 & \multicolumn{6}{c}{Amplitude (mmag)}\\
\hline
1 & & & 17.28 & 29.56 & & \\
2 & & & 23.61 & 20.24 & 15.12 & 11.59\\
3 & 14.74 & 16.25 & 16.07 & 17.96 & &\\
4 & & 20.83 & 8.83 & 10.14 & &\\
\hline
\end{tabular}
\end{center}
\end{table}

\subsubsection[]{Frequency determination test}
\label{robust}

To support our final frequency determination we performed a robustness test. Synthetic light curves were generated using frequencies, amplitudes and phases of the light curve solutions with Gaussian random noise added. The applied noise levels were scaled from 1 to 4 times the residual scatters ($\sigma$) of the light curve solutions. Independent frequency analyses were carried out for each dataset (1) to check how accurately we can get back the input frequencies from the noisy datasets and (2) whether we can identify the same frequencies that were found to be significant in the original datasets.

Testing our first subset we faced a $\pm$1--2\,c/d alias problem as we raised the noise level, especially in the case of frequencies with lower amplitudes. Frequencies were found rather far from the input values. At the 4\,$\sigma$ level we could identify 2 frequencies (99.25, 103.79\,c/d) out of the 4 found to be significant earlier. In the case of our second and third subsets, the amplitudes were rather high and we found all 4 of the frequencies identified before. Adding noise to the fourth subset caused alias problems from the 2.5\,$\sigma$ level upward and the lowest amplitude peaks disappeared. We could find only 4 frequencies (86.99, 90.10, 98.79 and 104.03\,c/d) within the most noisy time string.

Table~\ref{ref.nights.freq2} shows the results of our robustness test with the finally accepted frequencies. Peaks at $\sim$104 and 99\,c/d were found in all subsets. Other four frequencies (at $\sim$69, 87, 111 and 139\,c/d) were rather high amplitude peaks in one or two segments. These 6 frequencies were determined unambiguously by the analyses of subsets and perhaps characterize the pulsation of KUV 02464+3239 within some amplitude limit. 

As can be seen in Fig.~\ref{FT.ref.nights} and in Table~\ref{ref.nights.freq2}, remarkable amplitude variations occur from one subset to another. A signal at $\sim$87\,c/d became dominant in December while the amplitude of the peak at 104\,c/d decreased by 66\%. The peak at 99\,c/d first increased its amplitude by 37\% then decreased by 63\%.

\begin{figure}
\begin{center}
\includegraphics[width=8.5cm]{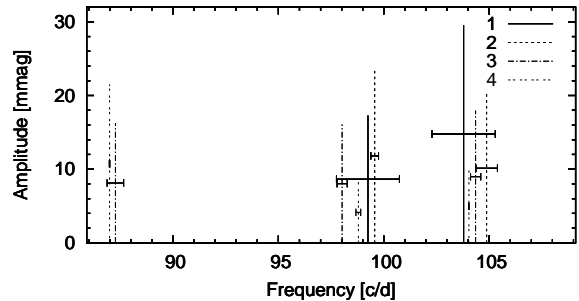}
\end{center}
\caption{Frequecies out of the 6 accepted ones found at least in two subsets. We calculated the width of error bars by the frequency range within the peaks were found during our robustness test (see Sect.~\ref{robust} for details).}
\label{scatter}
\end{figure}

We represent the frequencies which were found at least in two subsets together in Fig.~\ref{scatter}. The groups of frequencies are well-separated, there is no problem with overlapping sidelobes. The error bars give the uncertainty of our findings on a certain value in a subset. Since they overlap, it is clear that the two peaks around 87\,c/d can have different values because of the uncertainties. The error bars also overlap in the case of peaks at $\sim$104\,c/d. The four peaks around 99\,c/d represent an interesting case. Since the three well-determined peaks are more widely separated than the error bars, we can say that they are distinct. The reasons could be a frequency change on a short time scale (from one subset to the other -- within one month). Another explanation is that they are independent modes that are always excited and their amplitudes change from subset to subset. Alternatively, they might be components of an unresolved rotationally split mode. A comparative analysis on a longer time base could clear up the real situation (if it is a regular behaviour). At this early stage of the interpretation of KUV 02464+3239 we mention this behaviour but we give a general solution based on the whole dataset. 

\subsection[]{Analysis of the whole dataset}
\label{whole.lc.analysis}

The long time base of the whole dataset gives more precise frequency values if we assume the frequencies are stable over the time base. In this case the short-term variations in amplitudes are obscured but the frequencies are well-determined. 

\begin{figure}
\begin{center}
\includegraphics[width=9.0cm]{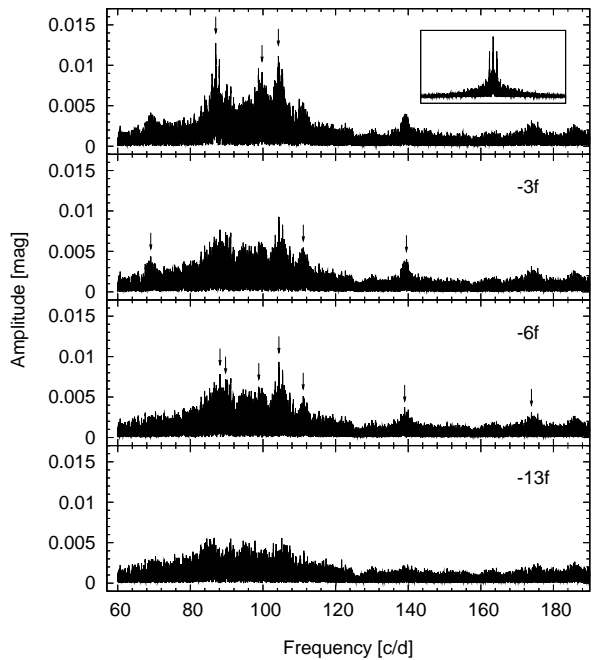}
\end{center}
\caption{Frequency analysis of the whole dataset. The panels show the successive pre-whitening steps, the window function is given in the insert. The 6 finally accepted frequencies are marked in the first and second panels. The third panel shows the residual spectrum with definite signals remained. The spectrum after pre-whitening with 13 frequencies can be seen on the last panel. It is still not white noise.}
\label{whitening}
\end{figure}

The successive pre-whitening steps of the whole dataset can be seen in the panels of Fig.~\ref{whitening}. Table~\ref{whole.freq} shows the values of frequencies determined from the whole dataset. Standard deviations are calculated by MC simulations. Peaks at $f_1$\,--\,$f_6$ are marked in the first and second panels of Fig.~\ref{whitening} (upper segment of Table~\ref{whole.freq}). They represent the 6 frequencies that come out of the final analysis. The residual spectrum in the third panel is definitely not white noise. With further analysis, we found 7 other frequencies ($\rmn{f}_7$\,--\,$\rmn{f}_{13}$) in the residual spectrum and present them in the lower segment of Table~\ref{whole.freq}. The last panel of Fig.~\ref{whitening} shows the FT after pre-whitening with 13 frequencies.

Amongst $\rmn{f}_7$, $\rmn{f}_8$ and $\rmn{f}_9$, additional pulsation modes could be present, but we cannot say which ones are real modes. $\rmn{f}_7$ is near to the $+$1\,c/d alias of the $f_2$ peak obtained in subset 3. There is no sign of $\rmn{f}_8$ in any subsets. $\rmn{f}_9$ can be a remainder of the group of four peaks near $f_3$ obtained by the subsets. The two higher frequencies in Table~\ref{ref.nights.freq2} (99.25, 99.57\,c/d) appear as $f_3$ and the two lower frequencies (98.02, 98.80\,c/d) can produce $\rmn{f}_9$ in the whole dataset. The double structure around 139\,c/d might be a sign of rotational splitting: $\delta f$ = 0.5\,c/d = 5.79\,$\mu$Hz. Assuming that these peaks belong to high-overtone ($k\gg1$) $l=1$, $m=-1,0$ or $m=0,1$ modes the rotation period of the star would be $\sim$1\,d. Considering the peaks at 104\,c/d, their frequency separation is only $\delta f$ = 0.13\,c/d = 1.5\,$\mu$Hz which corresponds to a $\sim$3.8\,d rotation period under the assumptions above. However, regarding the large amplitude of the closely spaced peak, $\rmn{f}_{10}$ could also be an independently excited mode with another $l$ value. Determination of the frequency at 111.09\,c/d ($\rmn{f}_{11}$) was ambiguous because of aliasing. The peak at 173.96\,c/d is consistent with being the first harmonic of the dominant mode.

For asteroseismological investigations, we use only the 6-frequency solution supported by the analyses of subsets. We consider these frequencies as a set of normal modes. A dataset with better coverage and higher signal-to-noise ratio would enable us to determine additional real pulsation modes.

\begin{table}
\caption{Frequency, period and amplitude values of 13 frequencies derived by the whole dataset. Frequencies at $f_1$\,--\,$f_6$ are accepted as normal modes by the analyses and test of subsets. Frequency and amplitude values of $f_1$\,--\,$f_6$ were determined by fitting with these components only. Frequencies and amplitudes of $\rmn{f}_7$\,--\,$\rmn{f}_{13}$ were calculated by fitting with all the 13 frequencies. Amplitude values of $f_1$\,--\,$f_6$ for the 13 frequency solution are given in parentheses.}
\label{whole.freq}
\begin{center}
\begin{tabular}{lrlrrr}
\hline
 & \multicolumn{3}{c}{Frequency} & \multicolumn{1}{c}{Period} & \multicolumn{1}{c}{Ampl.}\\
 & \multicolumn{2}{c}{(c/d)} & \multicolumn{1}{c}{($\mu$Hz)} & \multicolumn{1}{c}{(s)} & \multicolumn{1}{c}{(mmag)}\\
\hline
$f_1$ & 69.1060 & $\pm$0.0003 & 799.838 & 1250.253 & 4.4 (4.4)\\
$f_2$ & 86.9879 & $\pm$0.0001 & 1006.804 & 993.242 & 13.2 (14.4)\\
$f_3$ & 99.7516 & $\pm$0.0001 & 1154.532 & 866.151 & 9.5 (11.5)\\
$f_4$ & 104.2617 & $\pm$0.0001 & 1206.733 & 828.684 & 11.6 (11.4)\\
$f_5$ & 111.1095 & $\pm$0.0002 & 1285.989 & 777.611 & 5.5 (8.0)\\
$f_6$ & 139.5157 & $\pm$0.0003 & 1614.765 & 619.285 & 4.0 (3.9)\\
\hline
$\rmn{f}_7$ & 88.1611 & & 1020.383 & 980.024 & 8.4 \\
$\rmn{f}_8$ & 89.7100 & & 1038.311 & 963.103 & 7.3 \\
$\rmn{f}_9$ & 98.8533 & & 1144.136 & 874.022 & 6.5 \\
$\rmn{f}_{10}$ & 104.3915 & & 1208.235 & 827.654 & 10.2 \\
$\rmn{f}_{11}$ & 111.0865 & & 1285.724 & 777.772 & 5.8 \\
$\rmn{f}_{12}$ & 139.0162 & & 1608.983 & 621.510 & 3.7 \\
$\rmn{f}_{13}$ & 173.9642 & & 2013.474 & 496.654 & 3.2 \\
\hline
\end{tabular}
\end{center}
\end{table}

\section[]{Tests on amplitude variations}
\label{Ampl.var.}

The Fourier analysis method presumes stationary amplitude and frequency values while the behaviour of KUV 02464+3239 seems to be different. We do not know why the observed amplitudes of the star vary so much. To provide constraints for possible explanations, we performed tests for the amplitude variability.

We generated synthetic light curves corresponding to the whole dataset using the frequency, amplitude and phase values of either the 6 accepted frequencies or with the additional closely spaced frequencies around the 87\,c/d dominant mode. In every case, we analysed the four subsets by nonlinear least-squares fitting using the original frequency values as initial parameters. 

As our first test ($a$) we generated a synthetic light curve using only 6 frequencies and we did not add noise to the dataset. The analyses of subsets did not show noticeable amplitude variations from one subset to another. This result shows that beating of the 6 modes cannot explain the apparent amplitude variations. In our second test ($b$) we added Gaussian random noise to the synthetic light curve using the residual scatter of the whole light curve's 13-frequency solution. Our analysis showed that noise had a slight influence on the amplitudes (mainly for the lower amplitude modes) and cannot be responsible for the large observed variability. In our subsequent tests ($c$ -- $h$) we generated the synthetic light curves using 6+2 frequencies where the two additional frequencies were close to the dominant mode ($\delta f$ = $\pm$0.185\,c/d = 2.14\,$\mu$Hz). This separation was selected to be close to the resolution limit of the longest subset (subset 4) and this could be a reasonable value for rotational splitting. We set the amplitudes of the triplet's side components to be the half ($c$ -- $e$) and one tenth ($f$ -- $h$) of the 87\,c/d mode's amplitude. The phase differences were $\pm 22.5^{\circ}$ ($c$, $f$), $\pm45^{\circ}$ ($d$, $g$) and $\pm90^{\circ}$ ($e$, $h$) with respect to the central peak's phase at the initial epoch (JD\,2\,454\,014). With these tests we simulated the effects on amplitudes of closely spaced frequencies which might remain unresolved in the subsets. Such peaks could be detected by the analysis of the whole light curve if they had large enough amplitudes so as not to vanish in the other peaks and the noise.

\begin{table}
\caption{Results of tests on amplitude variability of the 87\,c/d mode. At test cases $a$ -- $h$ we give relative amplitude differences with respect to the synthetic light curves' input value. See details in Sect.~\ref{Ampl.var.}.}
\label{amplvar}
\begin{center}
\begin{tabular}{lrrrrrrrr}
\hline
\multicolumn{1}{c}{Subset} & \multicolumn{8}{c}{Test cases}\\[1.5mm]
 & \multicolumn{1}{c}{$a$} & \multicolumn{1}{c}{$b$} & \multicolumn{1}{c}{$c$} & \multicolumn{1}{c}{$d$} & \multicolumn{1}{c}{$e$} & \multicolumn{1}{c}{$f$} & \multicolumn{1}{c}{$g$} & \multicolumn{1}{c}{$h$}\\[2.0mm]
 & \multicolumn{8}{c}{Amplitude variation (\%)}\\
\hline
1 & 0 & +4 & +28 & +9 & -12 & -3 & +10 & -3\\
2 & 0 & -6 & +60 & +37 & -52 & +11 & +1 & -9\\
3 & 0 & +15 & -87 & -64 & -17 & -7 & -28 & +6\\
4 & 0 & -5 & +22 & +11 & -35 & +7 & +14 & -4\\ 
\hline
\end{tabular}
\end{center}
\end{table}

Table~\ref{amplvar} summarizes the test results on the amplitude variations of the dominant peak. In some cases we got large amplitude variations: disappearance and almost-disappearance of the peak in subset 3 (tests $c$ and $d$). In test $c$ we fitted a peak at 86.3\,c/d (the input value was 86.99\,c/d) with 1.5\,mmag amplitude. This means an amplitude decrease by 87\% if we accept the 86.3\,c/d peak as the representative of the original one. In test $d$ we got a 64\% decrease in amplitude. Triplet side components with smaller amplitudes caused smaller, but noticeable variations. The maximum was a 28\% decrease in the amplitude of the dominant peak (test $g$).

Amplitude variations were detected for other modes as well. The presence of noise and beating with triplet components can be the explanation since we generated the synthetic light curves using constant amplitude values. The average of the variations of the 5 modes is 15\%.

We conclude that the observed large amplitude variability can be simulated with a triplet assuming certain phase relations and relatively high amplitudes for the side components. However, these special phase relations and high amplitudes of side lobes to the main peak are rather improbable. A real amplitude variation seems to be plausible.

\section[]{Asteroseismology of KUV 02464+3239}
\label{seism}

Our main goal in asteroseismology is to find stellar models that match the observed properties of the star with the highest precision. The most important observed values are the periods of light curves. At first we are looking for models which produce pulsation periods with the lowest differences to the observed ones. We used the White Dwarf Evolution Code (WDEC) originally written by Martin Schwarzschild and modified by \citet{kutter1}, \citet{lamb1}, \citet{winget1}, \citet{kawaler1}, \citet{wood1}, \citet{bradley1}, \citet{montgomery1} and \citet{kim1} to generate equilibrium white dwarf models each with a given set of input parameters. For each model, we computed a set of periods based on the adiabatic equations of non-radial stellar oscillations \citep{unno1}. We computed grids of such models, varying the input parameters.

In WDEC we start with a hot ($\sim$100\,000\,K) polytrope model. The code evolves it down to the requested temperature and the model we finally get are thermally relaxed solution to the stellar structure equations. To get pulsation periods of a model we used the integrated form of the WDEC given by Metcalfe (2001), which includes a pulsation code.

We used the equation-of-state (EOS) tables of \citet{lamb2} in the core and the EOS tables of \citet{saumon1} in the envelope of the star. We used OPAL opacities updated by \citet{iglesias1}. The convection was treated within the Mixing Length Theory (MLT) according to \citet{bohm1} and with the $\alpha = 0.6$ parametrization suggested by model atmosphere calculations of \citet{bergeron2}. Precise value of $\alpha$ has very little effect on the pulsation periods of models (Bischoff-Kim, Montgomery \& Winget 2008b). Treatment of the hydrogen/helium transition zone was done by equilibrium diffusion calculations contrary to the treatment of the helium/carbon transition layer which was parametrized.

\subsection[]{The model grid}

As we discussed in Sect.~\ref{whole.lc.analysis} we used the 6-frequency light curve solution for the seismological analysis of KUV 02464+3239. Table~\ref{whole.freq} shows that we found modes only in the long period regime (between 619\,s and 1250\,s). \citet{kim3} shows that low order modes are especially sensitive to the mass of the hydrogen layer.

The best-fitting models have to fulfil two criteria: (1) to give back the period values closest to our observed values and (2) to provide physically acceptable solutions for effective temperature and $\rmn{log}\,g$. Finding solutions according to the first criterion we used the \verb"fitper" tool developed by Kim (2007). The program calculates the \textit{r.m.s.} values using the following equation:

%\[
\begin{equation}
\sigma_{r.m.s.} = \sqrt{\frac{\sum_{i=1}^{N} (P_i^{\rmn{calc}} - P_i^{\rmn{obs}})^2}{N}}
\label{equ1}
%\]
\end{equation}

\noindent where \textit{N} denotes the number of observed periods.

\begin{table*}
%\begin{minipage}[]{160mm}
\caption{Best-fitting solutions for KUV 02464+3239 in the mass range 0.525 -- 0.74\,$M_{\sun}$. We represent the results of former spectroscopic observations ($T_{\rmn{eff}}$, $\rmn{log\,} g$, \citealt{fontaine1}), the expected seismological mass derived by \citet{bradley3} and the observed pulsation periods as reference values. Models that have masses within 1\,$\sigma$ in $\rmn{log\,} g$ range are typeset in boldface.}
\label{best.fit}
%\begin{center}
\begin{tabular}{lccccccccccc}
\hline
\multicolumn{1}{l}{$M_*$/$M_{\sun}$, ($\rmn{log\,} g$)} & \multicolumn{1}{c}{$T_{\rmn{eff}}$\,(K)} & \multicolumn{1}{c}{-log\,$M_\rmn{H}$} & \multicolumn{1}{c}{$X_\rmn{o}$} & \multicolumn{1}{c}{$X_{\rmn{fm}}$} & \multicolumn{6}{c}{Model periods in seconds ($l$,$k$)} & \multicolumn{1}{c}{$\sigma_{r.m.s.}$\,(s)}\\
\hline
0.525, (7.85) & 11\,400 & 6.9 & 0.7 & 0.5 & 620.4 & 777.1 & 830.3 & 865.4 & 992.3 & 1250.7 & 0.95\\
 & & & & & (1,8) & (1,11) & (1,12) & (2,24) & (1,15) & (1,20) & \\[1.5mm]
0.525, (7.85) & 11\,600 & 7.2 & 0.5 & 0.4 & 620.3 & 776.3 & 830.7 & 866.0 & 992.6 & 1251.0 & 1.12\\
 & & & & & (2,16) & (1,11) & (1,12) & (2,24) & (1,15) & (1,20) & \\[1.5mm]
0.575, (7.97) & 11\,000 & 6.2 & 0.7 & 0.5 & 619.8 & 777.4 & 831.0 & 868.7 & 994.0 & 1250.5 & 1.45\\
 & & & & & (2,17) & (1,12) & (1,13) & (2,25) & (1,16) & (1,21) & \\[1.5mm]
0.585, (7.99) & 11\,000 & 6.4 & 0.5 & 0.4 & 617.9 & 775.4 & 827.1 & 866.9 & 991.6 & 1250.5 & 1.44\\
 & & & & & (2,17) & (1,12) & (1,13) & (2,25) & (1,16) & (1,21) & \\[1.5mm]
\textbf{0.615, (8.03)} & \textbf{11\,800} & \textbf{4.0} & \textbf{0.7} & \textbf{0.3} & \textbf{616.4} & \textbf{777.4} & \textbf{828.8} & \textbf{865.6} & \textbf{991.1} & \textbf{1249.7} & \textbf{1.51}\\
 & & & & & \textbf{(1,12)} & \textbf{(1,16)} & \textbf{(2,31)} & \textbf{(1,18)} & \textbf{(1,21)} & \textbf{(2,48)} & \\[1.5mm]
\textbf{0.625, (8.04)} & \textbf{11\,000} & \textbf{7.4} & \textbf{0.5} & \textbf{0.2} & \textbf{620.6} & \textbf{780.0} & \textbf{827.4} & \textbf{866.8} & \textbf{994.1} & \textbf{1252.1} & \textbf{1.50}\\
 & & & & & \textbf{(1,9)} & \textbf{(1,12)} & \textbf{(2,24)} & \textbf{(1,14)} & \textbf{(2,29)} & \textbf{(1,21)} & \\[1.5mm]
\textbf{0.645, (8.07)} & \textbf{11\,400} & \textbf{5.2} & \textbf{0.9} & \textbf{0.2} & \textbf{618.1} & \textbf{779.7} & \textbf{828.2} & \textbf{865.7} & \textbf{991.9} & \textbf{1251.8} & \textbf{1.33}\\
 & & & & & \textbf{(2,21)} & \textbf{(1,15)} & \textbf{(1,16)} & \textbf{(1,17)} & \textbf{(1,20)} & \textbf{(2,45)} & \\[1.5mm]
\textbf{0.650, (8.08)} & \textbf{11\,800} & \textbf{4.6} & \textbf{0.6} & \textbf{0.1} & \textbf{620.3} & \textbf{776.6} & \textbf{827.1} & \textbf{866.2} & \textbf{992.8} & \textbf{1249.6} & \textbf{0.93}\\
 & & & & & \textbf{(1,12)} & \textbf{(2,29)} & \textbf{(1,17)} & \textbf{(1,18)} & \textbf{(1,21)} & \textbf{(2,48)} & \\[1.5mm]
\textbf{0.680, (8.13)} & \textbf{11\,800} & \textbf{5.0} & \textbf{0.5} & \textbf{0.1} & \textbf{618.8} & \textbf{778.1} & \textbf{826.6} & \textbf{865.2} & \textbf{992.1} & \textbf{1251.9} & \textbf{1.26}\\
 & & & & & \textbf{(1,12)} & \textbf{(2,29)} & \textbf{(1,17)} & \textbf{(1,18)} & \textbf{(1,21)} & \textbf{(2,48)} & \\[1.5mm]
\textbf{0.685, (8.14)} & \textbf{11\,400} & \textbf{4.8} & \textbf{0.6} & \textbf{0.4} & \textbf{618.7} & \textbf{778.6} & \textbf{826.4} & \textbf{866.4} & \textbf{994.1} & \textbf{1249.9} & \textbf{1.12}\\
 & & & & & \textbf{(1,12)} & \textbf{(1,16)} & \textbf{(1,17)} & \textbf{(2,32)} & \textbf{(2,37)} & \textbf{(2,47)} & \\[1.5mm]
0.725, (8.2) & 11\,800 & 5.8 & 0.7 & 0.4 & 620.4 & 778.5 & 828.4 & 865.6 & 995.3 & 1249.5 & 1.11\\
 & & & & & (1,12) & (1,16) & (1,17) & (1,18) & (1,21) & (1,27) & \\[1.5mm] 
0.725, (8.2) & 11\,200 & 5.4 & 0.8 & 0.4 & 616.9 & 778.5 & 828.9 & 865.2 & 992.7 & 1251.6 & 1.26\\
 & & & & & (1,12) & (1,16) & (1,17) & (1,18) & (2,37) & (2,47) & \\[1.5mm]
0.740, (8.23) & 11\,800 & 6.0 & 0.6 & 0.4 & 620.0 & 778.9 & 828.1 & 867.0 & 991.3 & 1250.7 & 1.09\\
 & & & & & (1,12) & (1,16) & (1,17) & (1,18) & (1,21) & (1,27) & \\[1.5mm]
Reference values: & & & & & & & & & \\[1.5mm]
0.65, (8.08) & 11\,290 & & & & 619.3 & 777.6 & 828.7 & 866.2 & 993.2 & 1250.3 & \\
\hline
\end{tabular}
%\end{center}
%\end{minipage}
\end{table*}

We built our model grid varying 5 input parameters of the WDEC: $T_{\rmn{eff}}$, $M_*$, $M_{\rmn{H}}$, $X_\rmn{o}$ (central oxygen abundance) and $X_{\rmn{fm}}$ (the fractional mass point where the oxygen abundance starts dropping). For our first scan we fixed the mass of the helium layer at $M_{\rmn{He}} = 10^{-2}\,M_*$. We built our grid and searched for acceptable solutions in the parameter space determined by spectroscopic results on the star. \citet{fontaine1} derived $T_{\rmn{eff}} = 11\,290$\,K and $\rmn{log\,} g = 8.08$ values for KUV 02464+3239 with the estimated flux calibration uncertainties of $\pm200$\,K and 0.05\,dex in $\rmn{log\,} g$ \citep{fontaine2}. However, this uncertainty estimate does not include contributions from different spectra, modeling uncertainties, and different fitting procedures. We make some allowance for this by covering a range of $\pm500$\,K in $T_{\rmn{eff}}$ and $\pm0.1$\,dex in $\rmn{log\,} g$. According to this effective temperature value, our grid covers the range 10\,800 to 11\,800\,K ($\sim$11\,290\,$\rmn{K}\pm$2.5$\sigma$). Estimating the mass range we have to cover, we used the seismological masses determined for DAs by \citet{bradley3}. According to his results the mass of a DA white dwarf with $\rmn{log\,} g = 8.08$ is about $0.65\,M_{\sun}$. Our grid covers the $M_* = 0.525 - 0.74\,M_{\sun}$ mass range ($\rmn{log\,} g = \,\sim7.9 - 8.2$) which means that we investigated model solutions at $\rmn{log\,} g = 8.08\pm$2$\sigma$. We varied the mass of the hydrogen layer between $10^{-4} - 10^{-8}\,M_*$. The core parameters were changed between $X_\rmn{o} = 0.5 - 0.9$ and $X_{\rmn{fm}} = 0.1 - 0.5$ taking into account the carbon-oxygen profiles derived by \citet{salaris1}. Step sizes were 200\,K ($T_{\rmn{eff}}$), 0.005\,$M_{\sun}$ ($M_*$), $10^{-0.2}\,M_*$ ($M_{\rmn{H}}$) and 0.1 ($X_\rmn{o}$ and $X_{\rmn{fm}}$).

\subsection[]{Best-fit models}

While running the WDEC we allowed the modes to be $l = 1$ or 2. Since there are no previous asteroseismological analyses on this star or other contrainst on the $l$ values of modes we let all 6 modes be $l = 1$ or 2 for the fitting procedure. The variations in pulsation amplitudes rule out a meaningful application of different weights on the amplitudes \citep{castanheira1}.

Assuming better visibility of $l = 1$ modes over $l = 2$ modes (see e.g. \citealt{castanheira1} and references therein), we prefer models that gave at least 3 $l = 1$ solutions with low $\sigma_{r.m.s.}$ values. We summarize the parameters of the best-fitting models in Table~\ref{best.fit}. It shows the model parameters, calculated pulsation periods with their $l$ and $k$ values and the corresponding $\sigma_{r.m.s.}$ value. We find that the mode identifications of the best-fitting models are typically quite different from model to model. This is due to the fact that over the span of 400\,K, the periods of the relatively high overtone modes observed can change by about 40\,s. As a result, the cooling of models for a given structure causes different $l=1$ or 2 modes to fit a given observed period best. Note that since our current models use about 400 zones, the uncertainty of the model periods is $\sim$1\,s \citep{brassard1}, therefore, differences in periods below 1\,s are within the `noise' of our modeling accuracy. The presence of shorter period modes could give stronger constraints on the stellar structure, resulting in fewer solutions with low $\sigma_{r.m.s.}$. However, within the $\sim$1\,s uncertainty of our modeling we can map the possible groups of models in the parameter space determined by spectroscopy.

\subsubsection[]{Results for the main stellar parameters}

\begin{figure}
\begin{center}
\includegraphics[width=9.2cm]{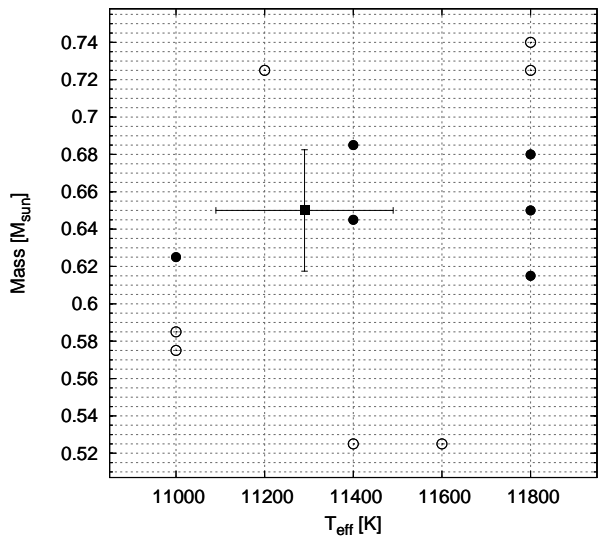}
\end{center}
\caption{The 13 models of Table~\ref{best.fit} in the $T_{\rmn{eff}}$ -- $M_*$ plane. Closed and open circles denote the models within and out of the 1\,$\sigma$ $\rmn{log\,} g$ range determined by spectroscopy. The black square denotes the spectroscopic solution with its uncertainties given by Fontaine et al. (2001, 2003). The grid with dashed lines corresponds to the step sizes of our model grid.}
\label{grid}
\end{figure}

Models with stellar masses within the 1\,$\sigma$ $\rmn{log\,} g$ range are typeset in boldface in Table~\ref{best.fit}. We found 6 models out of the 13 that fulfill this criterion. The best fit over all with $\sigma_{r.m.s.} = 0.93$\,s has $M_* = 0.65\,M_{\sun}$ which corresponds with the expected mass from the spectroscopic $\rmn{log\,} g$ value. Fig.~\ref{grid} shows the 13 selected models in the $T_{\rmn{eff}}$ -- $\rmn{log\,} g$ plane. Closed and open circles denote the models within and out of the 1\,$\sigma$ $\rmn{log\,} g$ range, respectively. The spectroscopic solution with the corresponding uncertainties determined by Fontaine et al. (2001, 2003) is also given in Fig.~\ref{grid}.

Assuming that the largest amplitude modes at 829, 866 and 993\,s are $l = 1$, we favour the models which give this solution. Within the 1\,$\sigma$ $\rmn{log\,} g$ range we find the 0.645, 0.650 and 0.680\,$M_{\sun}$ models in accordance with this criterion. Taking into account that $T_{\rmn{eff}} = 11\,290 \pm 200$\,K is expected also by spectroscopy, the 11\,800\,K solutions are over the 1\,$\sigma$ limit and seem to be too hot. This means that the 0.645\,$M_{\sun}$ model would be our best choice.

Table~\ref{best.fit} shows the hydrogen layer masses of our models are between $10^{-4} - 4\,\rmn{x}\,10^{-8}\,M_*$. Restricting ourselves to the 6 models selected by $\rmn{log\,} g$ we find that 5 of them have $M_\rmn{H} = 10^{-4} - 6\,\rmn{x}\,10^{-6}\,M_*$. We can give additional constraints on the hydrogen layer mass if we take into account only the three `favoured' models between 0.645 and 0.680\,$M_{\sun}$. In this case $M_\rmn{H} = 2.5\,\rmn{x}\,10^{-5} - 6.3\,\rmn{x}\,10^{-6}\,M_*$.

\subsubsection[]{Mode identification}

According to our selection criterion the models have mostly $l = 1$ solutions for the pulsation modes. Models with solely $l = 1$ modes exist only out of the 1\,$\sigma$ $\rmn{log\,} g$ range ($M_* = 0.725$ and $0.74\,M_{\sun}$).

Considering the full range of best-fit models, we cannot uniquely assign an $l$ value to any mode. At least one model has an $l = 2$ mode solution for each period. We cannot provide unique assignments for the overtone numbers either, however, the range of stellar and hydrogen layer masses of our models constrains the $k$ values.

In summary we can say that the observed pattern of modes can be decribed with high accuracy assuming mostly $l = 1$ modes. Considering the 3 `favoured' models the modes at 619 and 778\,s are always different $l$ values. The 1250\,s mode is always $l = 2$ for both these models.

\subsubsection[]{Varying the He layer mass}  

In our second scan, we varied the mass of the He layer in the grid from $10^{-2}$ to $10^{-3}\,M_*$ in 3 steps. This way we could check the sensitivity of our solutions to the thickness of the He layer. In this case we ran a more sophisticated version of the WDEC with core composition profiles based on stellar evolution calculations done by \citet{salaris1}. This grid covers the temperature and hydrogen layer mass range we used previously ($T_{\rmn{eff}} = 10\,800 - 11\,800$\,K and $M_\rmn{H} = 10^{-4} - 10^{-8}\,M_*$). We varied the stellar mass between $0.60 - 0.69\,M_{\sun}$. Step sizes were as well 200\,K ($T_{\rmn{eff}}$), 0.005\,$M_{\sun}$ ($M_*$) and $10^{-0.2}\,M_*$ ($M_{\rmn{H}}$).

According to our first criterion we searched for models with at least 3 $l = 1$ solutions and low $\sigma_{r.m.s.}$ values. We found two acceptable models at 0.61 and 0.635\,$M_{\sun}$. Both models have the same helium layer masses ($10^{-3}\,M_*$). However, none of them have all of the large amplitude modes as $l = 1$. The mode at 993\,s is always $l = 2$, which would imply a much larger physical amplitude for this mode since it has the largest light amplitude. These two models have hydrogen layers 
with masses of $2.5\,\rmn{x}\,10^{-6}$ and $10^{-6}\,M_*$. These values are at the thin side of the range given by the previously selected models. Although we cannot rule out helium layer masses thinner than $10^{-2}\,M_*$, our favouring the three largest amplitude modes being $l = 1$ implies that $M_{\rmn{He}} = 10^{-2}\,M_*$ is the preferred value.

\subsubsection[]{Seismological parallax}

According to the luminosity of a selected model we can give an estimation on the distance and parallax of the star. Unfortunately, there is no trigonometric parallax measurement for KUV 02464+3239. However, we can still predict what the parallax should be.

The first step is to derive the bolometric magnitude ($M_{\rmn{bol}}$) of the star:

%\[
\begin{equation}
M_{\rmn{bol}} = M_{\sun\,\rmn{bol}} - 2.5\,\rmn{log}\,(L/L_{\sun})\,,
%\]
\end{equation}

\noindent where $M_{\sun\,\rmn{bol}} = +4.75$ \citep{allen1} and $L_{\sun} = 3.854\,\rmn{x}\,10^{33}$\,ergs/s \citep{sackmann1}. $M_{\rmn{bol}}$ can be related to the absolute visual magnitude using the bolometric correction: $M_V = M_{\rmn{bol}} - \rmn{BC}$. Bergeron, Wesemael \& Beauchamp (1995a) performed colour-index and magnitude calculations using DA and DB model grids. According to Table 1 in \citet{bergeron1} $\rmn{BC} = -0.441$ and -0.611 at temperatures 11\,000 and 12\,000\,K, respectively. In view of the apparent visual magnitude $m_v = \rmn{16.07}$ and using the distance modulus formula we can derive the distance and the parallax of KUV 02464+3239.

The 6 models selected by $\rmn{log}\,g$ have $\rmn{log} \,L/L_{\sun} = -2.56$ to $-2.71$ where the most and least luminous models are at 0.615 and 0.625\,$M_{\sun}$, respectively. In these cases, the calculated seismological parallax values are between $13.5 - 15.4$\,mas (milliarc second) with a 14.6\,mas average. Restricting ourselves to the 3 `favoured' models with $\rmn{log}\,L/L_{\sun} = -2.6$ to $-2.66$, the parallax angles are 14.8, 14.2 and 14.7\,mas for the 0.645, 0.650 and 0.680\,$M_{\sun}$ models, respectively. These results imply a distance of $\sim$70\,pc for KUV 02464+3239.

\subsection[]{Mode trapping}

\begin{figure}
\begin{center}
\includegraphics[width=9.5cm]{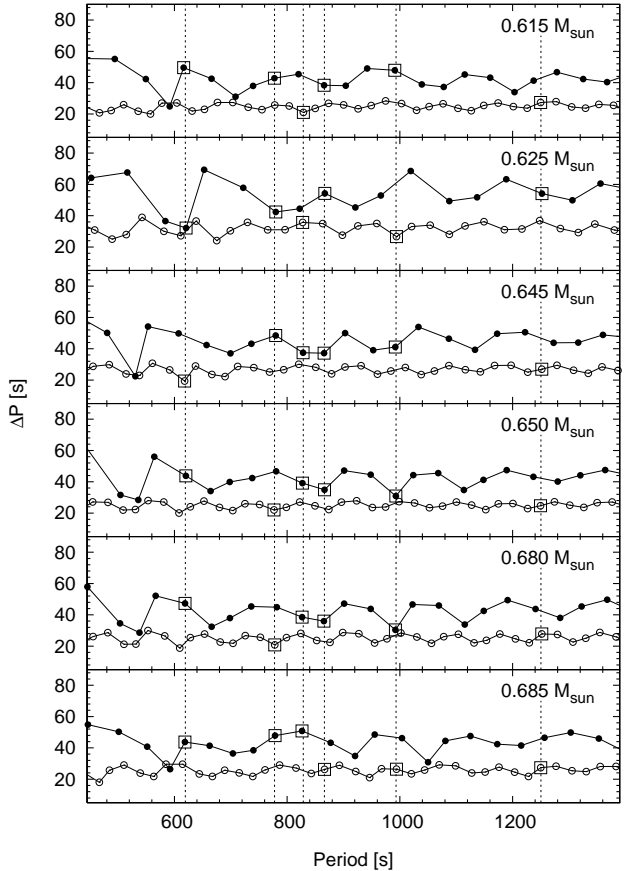}
\end{center}
\caption{Period spacing diagrams of the 6 selected models by $\rmn{log}\,g$. Filled and open circles denote the model periods with $l=1$ or $l=2$ solutions, respectively. Vertical dashed lines are drawn corresponding to the observed period values. Model periods found to be the closest of the observed ones are denoted with open squares.}
\label{periodsp}
\end{figure}

KUV 02464+3239 was mentioned by \citet{fontaine1} to be a photometric twin of another cool DAV, GD 154. In spite of the similarities between the shape of their light curves we find significant differences if we compare their pulsational properties. The Fourier spectrum of GD 154 is dominated by three normal modes and their harmonics \citep{pfeiffer1}. Subharmonics of the 1186\,s mode were also detected by \citet{robinson1}. In the case of KUV 02464+3239 even with single-site observations we could determine 6 modes and using the whole light curve, only the first harmonic of the dominant mode was found. Clear presence of further harmonics and subharmonics can be seen only by some parts of the light curve \citep{bognar1}.

The small number of GD 154's eigenmodes compared to other cool DAV's implies a very efficient mode selection mechanism -- possibly trapping of the modes by a very thin ($M_\rmn{H} \sim 10^{-10}\,M_*$) hydrogen layer \citep{pfeiffer1}. We can use our models on KUV 02464+3239 to test whether some of the modes are trapped.

One possible way to find trapped modes is to construct period spacing diagrams and search for minima. Mode trapping causes departures from the uniform period spacings of consecutive $k$ modes. Trapped modes have lower kinetic energies and as \citet{bradley5} showed, minima in period spacing diagrams correspond to minima in kinetic energies. Accordingly we made a simple statistical fit on the 6 models selected by $\rmn{log}\,g$. We collected how often we find a period near or at a minimum of a model's corresponding period spacing diagram. We found that the largest amplitude modes (at 829, 866 and 993\,s) frequently occur near or at period spacing minima and in these cases the $l = 1$ solution is preferred. The 1250\,s mode has not been found at a minimum in any of the cases. Fig.~\ref{periodsp} shows the period spacing diagrams of the 6 selected models. Filled and open circles denote the model periods with $l=1$ or $l=2$ solutions, respectively. Vertical dashed lines are drawn corresponding to the observed period values.

\section[]{Summary and conclusions}
\label{sum}

We have presented the analyses of our observations on the DA variable KUV 02464+3239.

Using analyses of data subsets we accepted 6 frequencies as normal modes of pulsation. With the analysis of the whole dataset we determined 7 additional frequencies as possible modes. Since remarkable amplitude variations (up to 30 -- 60\,\%) occured on a time scale of weeks, we performed tests for the possible sources of the amplitude variability. Neither beating of the 6 modes nor the effect of noise resulted in large enough variations. The test cases of closely spaced modes or rotational triplets revealed large amplitude variability, but only in unrealistic amplitude and phase relations. A real amplitude variation seem to be plausible.

The best asteroseismological models have $M_* = 0.645, 0.650$ and $0.680\,M_{\sun}$ and the mass of their hydrogen layer is between $2.5\,\rmn{x}\,10^{-5} - 6.3\,\rmn{x}\,10^{-6}\,M_*$. Although we cannot exclude thinner helium layers, $M_{\rmn{He}} = 10^{-2}\,M_*$ models reproduce the observed properties of the star better. Asteroseismological parallax values calculated by the luminosity of the 3 models are between 14.2 -- 14.8\,mas. Using period spacing diagrams we conclude that mode trapping can explain the high amplitude modes, but is not required.

The best way to discriminate between models would be to obtain observational constraints on the $l$ values of at least the large amplitude modes. It may be possible by time dependent UV spectra or chromatic amplitudes but due to the faintness of the star and its complicated pulsation spectra this would be a challenging task. Our results show the effectiveness of long, single-site observations. Multisite campaign(s) on KUV 02464+3239 would enable us to find additional pulsation modes and possibly signs of rotational splittings, and this would give further constraints on the stellar structure.

Resolving the closely spaced modes expected in the case of KUV 02464+3239 we could exclude the beating of modes as cause of the observed amplitude variability. Our tests on amplitudes indicate this possibility. This would mean that we see real energy content variations of the individual modes. In this case the observed variability is presumably due to non-linear processes like mode coupling and/or interaction between pulsation and convection. Finding the real physical explanation of amplitude variations is a key issue not only in the case of KUV 02464+3239, but also for other long-period white dwarf pulsators.

\section*{Acknowledgments}

The authors would like to thank the referee, S.~O. Kepler for his helpful comments and suggestions. This research was partly supported by HSO project No.\,98022.

\bsp

\label{lastpage}

\end{document}